# Tunable Plasmonic Absorption in Metal–Dielectric Multilayers via FDTD Simulations and an Explainable Machine Learning Approach

*†*Emmanuel A. Bamidele*

*Computer Science Department, Georgia Institute of Technology, USA*

*Materials Science and Engineering Program, University of Colorado Boulder, USA*

ABSTRACT

Plasmonic devices, fundamental to modern nanophotonics, exploit resonant interactions between light and free electrons in metals to achieve enhanced light trapping and electromagnetic field confinement. However, modeling their complex, nonlinear optical responses remain computationally intensive. In this work, we combine finite-difference time-domain (FDTD) simulations with machine learning (ML) to simulate and predict absorbed power behavior in multilayer plasmonic stacks composed of $SiO_2$, gold (Au), silver (Ag), and indium tin oxide (ITO). By varying Au and Ag thicknesses (10–50 nm) across a spectral range of 300–1500 nm, we generate spatial absorption maps and integrated power metrics from full-wave solutions to Maxwell's equations. A multilayer perceptron models global absorption behavior with a mean absolute error (MAE) of 0.0953, while a convolutional neural network predicts spatial absorption distributions with an MAE of 0.0101. SHapley Additive exPlanations identify plasmonic layer thickness and excitation wavelength as dominant contributors to absorption, which peaks between 450 and 850 nm. Gold demonstrates broader and more sustained absorption compared to silver, although both metals exhibit reduced efficiency outside the resonance window. This integrated FDTD–ML framework offers a fast, explainable, and accurate approach for investigating tunable plasmonic behavior in multilayer systems, with applications in optical sensing, photovoltaics, and nanophotonic device design.

INTRODUCTION

Plasmonic nanostructures have emerged as critical components in nanophotonic technologies, with their extraordinary ability to confine and manipulate light at subwavelength scales through resonant interactions between incident electromagnetic waves and free electrons in metals[1]. These interactions give rise to surface plasmon resonances (SPRs), which enhance local electromagnetic fields, enabling a wide array of applications spanning high-efficiency photovoltaics, ultrasensitive biochemical sensing, photothermal therapy, and advanced metamaterials. By tailoring the geometry and material composition of plasmonic devices, researchers have unlocked new ways to control light propagation, energy conversion, and heat generation, providing transformative solutions to challenges in energy, healthcare, and information technologies[2–4].

Despite the extensive progress made in the design and fabrication of plasmonic devices, a comprehensive understanding and optimization of their optical performance remain formidable tasks[5]. The non-linear nature of plasmonic resonances, coupled with the multidimensional design space involving material selection, layer thicknesses, and geometric configurations, complicates the prediction of their optical responses. Conventional design approaches, often relying on simplified analytical models, scientific intuition, or iterative electromagnetic simulations, have proven effective for relatively simple structures, such as metallic nanoparticles or core-shell geometries[6,7]. However, as device architectures grow more complex, particularly in multilayer configurations where multiple materials interact, these methods become computationally prohibitive and time-intensive.

To address these challenges, computational electromagnetic methods, such as the finite-difference time-domain (FDTD) technique, have become useful tools for solving Maxwell's equations and modeling light–matter interactions in plasmonic systems. The FDTD method discretizes space and time, providing high-resolution insights into field distributions, scattering cross-sections, and absorbed power densities. However, comprehensive exploration of the vast design parameter space using FDTD simulations alone is constrained by computational overhead, especially when optimizing multi-layer configurations for specific application-driven performance metrics. Traditional optimization algorithms, such as gradient-based topology optimization or evolutionary methods like genetic algorithms, often require iterative simulations and fail to efficiently explore large, multi-constrained design spaces. Furthermore, these methods struggle to capture the underlying non-linear dependencies between structural parameters and optical outcomes, leading to suboptimal solutions or prolonged convergence times.

In recent years, the emergence of machine learning (ML) has catalyzed a transformative shift in materials design, redefining how advanced materials are developed, optimized, and applied across diverse fields [8]. By leveraging data-driven models, ML enables accelerated discovery, property prediction, and structural optimization, particularly in complex domains like metamaterials[9] and nanostructures [10]. One of the most notable areas of impact is photonic/nanophotonic design, where ML techniques have revolutionized the engineering of optical materials and devices[11,12]. From inverse design of photonic crystals to the optimization of optical waveguides and resonators[13–15], ML-based approaches allow for the rapid identification of novel architectures[11], improving device efficiency, miniaturization, and functionality beyond conventional limitations. Deep learning algorithms - particularly convolutional neural networks (CNNs) - have proven adept at capturing spatial correlations and learning complex, high-dimensional mappings across diverse scientific applications[16]. By leveraging substantial datasets derived from high-fidelity simulations, CNNs can discern intricate relationships between device configurations and their optical responses without necessitating exhaustive parametric sweeps. This capability to generalize across parameter spaces and predict absorbed power distributions with high accuracy signals new opportunities for efficient plasmonic device optimization.

In this work, we present an integrated framework that combines finite-difference time-domain (FDTD) simulations with two complementary ML models, including (a) CNN and (b) multilayer perceptron (MLP). The CNN focuses on the spatial distribution of absorbed power density in multilayer plasmonic structures, while the MLP targets absorption metrics. We model realistic device stacks incorporating silicon dioxide ($SiO_2$), gold (Au), and indium tin oxide (ITO), examining how variations in Au-layer thickness and excitation wavelength (300–1500 nm) impact plasmonic resonances. By encoding material properties, geometrical parameters, and spectral data as input features, the CNN learns the mapping from structural configurations to high-dimensional power-density distributions, whereas the MLP efficiently predicts aggregate absorbed flux and power.

This dual-model approach not only accelerates the search for optimal device designs but also sharply reduces computational costs relative to traditional methods. In addition, we incorporate SHapley Additive exPlanations (SHAP) to identify the key factors dictating absorption performance, elucidating how thickness, material composition, and wavelength work in tandem to govern plasmonic efficiency. Our results highlight the promise of this data-driven methodology for bridging the gap between theoretical simulations and real-world applications, particularly in areas such as energy harvesting, optical sensing, and photothermal therapy-domains where effective light–matter interaction is paramount.

MAIN

The multi-layer structure consists of $SiO_2$, Au (or Ag), and ITO layers. The thickness of the gold and silver layers, the primary plasmonic component, ranges from 10 to 50 nm to explore resonant coupling. $SiO_2$ and ITO thicknesses are fixed at 500nm and 200nm respectively. The multi-layer stack and the simulation cell structure are represented in **Figure 1**.

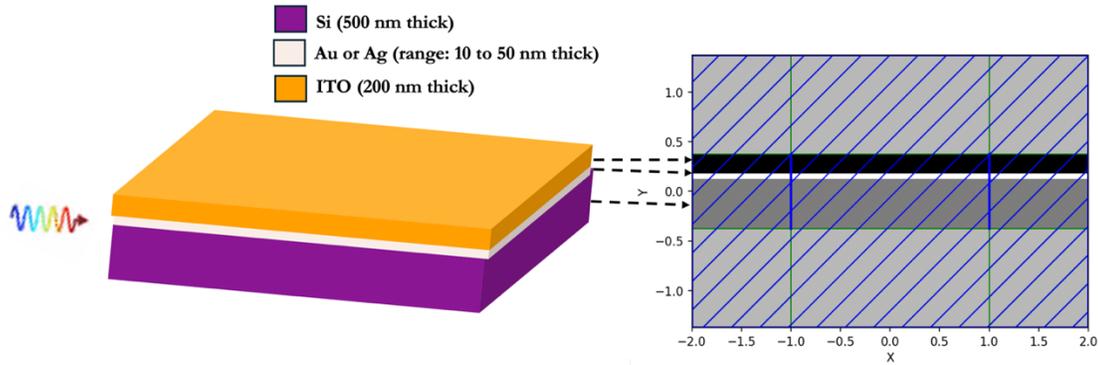

*Figure 1*: Left: Multilayer structure of $SiO_2$, Au/Ag and ITO with 500, 200 and 10-50nm thicknesses respectively. Right: Cell Structure of the simulation setup with a 4 x 2.75 μm cell structure.

**Wavelength-Dependent Optical Absorption and Plasmonic Resonances**

**Figure 2** shows the measured and simulated optical absorption spectra of the Au/Ag multilayer structure as a function of wavelength. A pronounced absorption peak is observed, indicative of a plasmonic resonance – a phenomenon arising from collective oscillations of conduction electrons at the metal–dielectric interface – whose frequency and intensity are strongly influenced by the dielectric environment [17,18]. The resonance wavelength is determined by the effective refractive index of the surrounding media and the plasmonic layer geometry: a higher surrounding index or a larger effective layer thickness leads to a red-shift of the plasmon peak[17]. This behavior is consistent with classical plasmonic theory, which predicts that the SPR condition depends sensitively on the dielectric constant of the adjacent medium [17]. For example, embedding the multilayer in a higher-index medium or adding a high-index capping layer increases the effective permittivity seen by the plasmon, shifting the resonance to longer wavelengths (lower energy) [17]. Such dielectric tuning of plasmonic absorption has been widely reported in nanoparticle systems and is utilized in refractive-index sensing applications [17]. The broadening and amplitude of the absorption band are dictated by the intrinsic optical response of the metals. Noble metals like Au and Ag exhibit a negative real permittivity in the visible range due to the free-electron (Drude) response of the conduction band, while interband transitions contribute additional dispersion and loss at higher energies (ultraviolet and blue visible) [18,19].

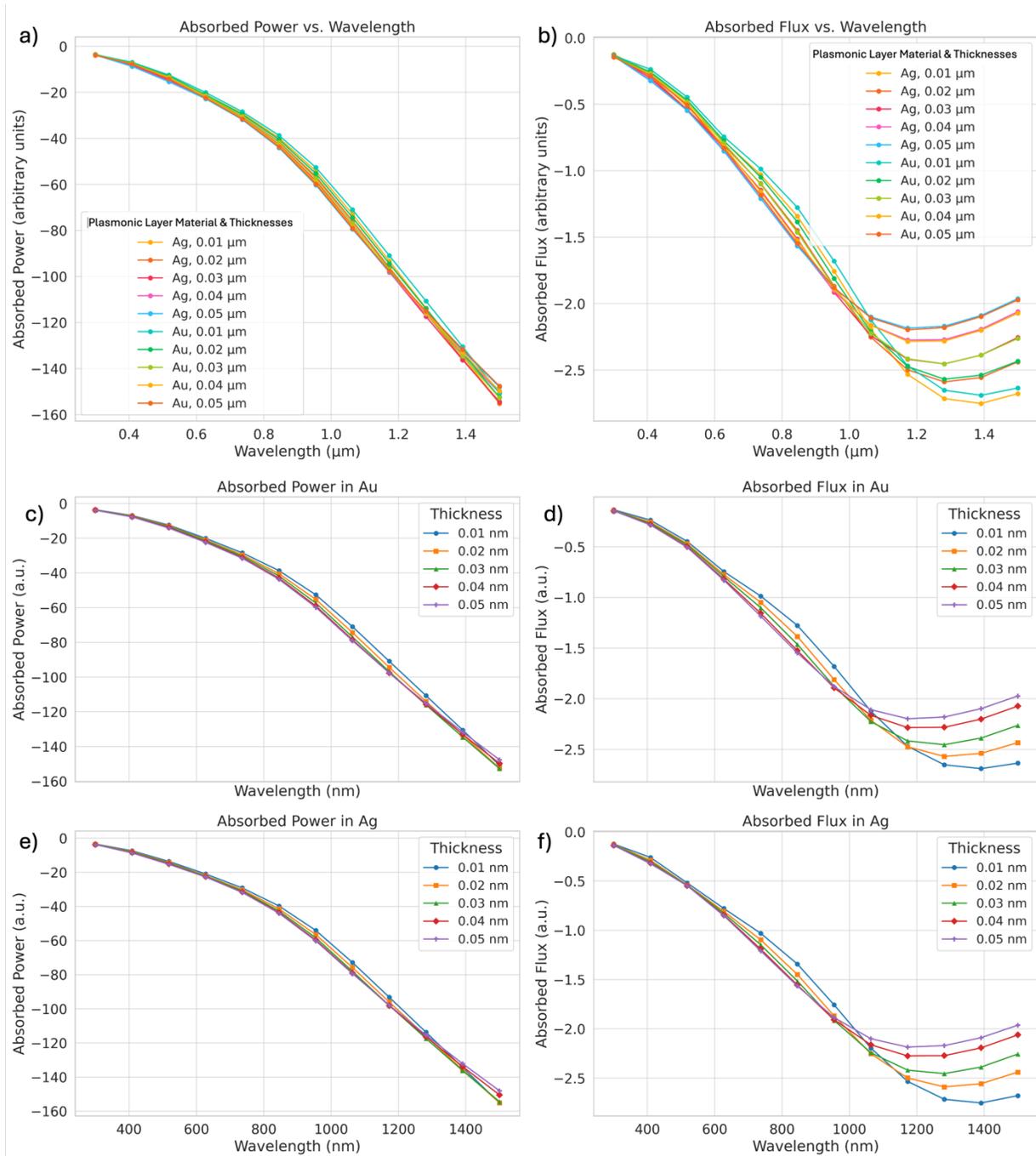

*Figure 2*: *Absorbed power and absorbed flux across Au and Ag plasmonic layers and their thicknesses. (a, b) Overall absorbed power and flux versus wavelength for plasmonic layers of both materials (Ag, Au) and thicknesses. (c, d) Absorbed power and flux specifically in Au layers at various thicknesses, showing . (e, f) Absorbed power and flux in Ag layers across different thicknesses Two key inflection points are evident near 450 nm and 850 nm, corresponding to changes in interband damping and loss of plasmonic resonance conditions, respectively. The flux curves at longer wavelengths (850–900 nm) underscore the impact of reduced layer thickness on plasmonic absorption efficiency.*

According to the Drude model, the frequency-dependent dielectric function of a metal can be expressed as:

$$\epsilon(\omega) = \epsilon_\infty - \frac{\omega_p^2}{\omega(\omega + i\gamma)}$$

where $\omega_p$ is the plasma frequency, representing the oscillation of free electrons. $\gamma$ is the damping constant that accounts for energy loss due to electron scattering and $\epsilon\infty$ is the high-frequency dielectric constant. For bulk Au, $\omega_p$ corresponds to an energy of approximately 9 eV, so in the visible range ($\hbar\omega \sim$ 1–3 eV) the real part of $\epsilon(\omega)$ is negative, enabling the existence of surface plasmons at metal–dielectric interfaces[19]. The resonance condition for an interface plasmon is approximately $Re[\epsilon(\omega)] \approx -\epsilon_d$, where $\epsilon_d$ is the dielectric constant of the adjacent medium [18]. In our multilayer system, at the absorption peak (~600 nm for the Au-based structure), the real part of the effective permittivity of the metallic layer approaches this condition, allowing strong field confinement and energy coupling into the metal. We note that a simple Drude model is insufficient to fully describe Au's optical response above ~2.4 eV (520 nm); a Lorentz term (or additional oscillators) must be included to account for interband transitions of bound electrons [19]. These interband transitions in Au cause a rise in absorption toward shorter wavelengths (the shoulder of the spectra below 500 nm in **Figure 2**), whereas at longer wavelengths beyond the plasmon band the metal becomes more reflective and the absorption drops off, as expected when $\omega < \omega_p$ in the Drude picture. Overall, the good agreement between our simulations and the Drude–Lorentz theory indicates that the observed spectral features, a dominant plasmonic absorption band and increased absorption in the blue, originate from the frequency-dependent dielectric function of the Au/Ag layers. This provides a solid theoretical foundation for understanding and tuning the multilayer's optical response.

**Thickness-Dependent Absorption and Spatial Distribution**

We next investigated how the optical absorption is affected by the thickness of the metal layers in the multilayer. **Figure 3** presents the simulated absorption spectra for varying Au layer thickness (with other parameters fixed). As the Au thickness increases from ultrathin (~5 nm) to moderate (~30 nm), the peak absorption at the plasmon resonance initially rises, reflecting the greater volume of metal available to absorb energy. We find that there is an optimal thickness (around 20 nm in our configuration) at which the absorption is maximized, exceeding 80% at the resonance. Beyond this point, further increases in thickness yield diminishing returns: the absorption peak amplitude saturates or even slightly decreases for very thick films. This trend can be explained by the interplay of field penetration and reflectivity. For thin films, the evanescent plasmon field penetrates through the metal and a significant portion of the incident light reaches the second interface, whereas for thick films the metal layer reflects more light and the plasmon field is mostly confined to the front interface. There exists an optimum thickness at which the impedance of the plasmonic layer is best matched to free space, maximizing the absorption of incident light (similar to the concept of a Salisbury absorber or metamaterial perfect absorber). Notably, when the metal layer is extremely thin (<5 nm), the absorption drops sharply in our simulations. In this regime, the film would be discontinuous in practice, forming isolated nano-islands, and thus supports weaker, blue-shifted plasmonic modes. Indeed, experiments have shown that a 2 nm evaporated Au film behaves as an ensemble of small nanoparticles with an absorption peak around 600 nm, in contrast to a continuous 10–20 nm film which exhibits the SPR near 520 nm characteristic of bulk gold [20]. Our simulations (which assume continuous films) do not explicitly capture the percolation transition, but the trend of reduced and shifted absorption for the thinnest layers is consistent with those observations [20]. These results underscore the importance of thickness control in tuning plasmonic multilayer absorbers.

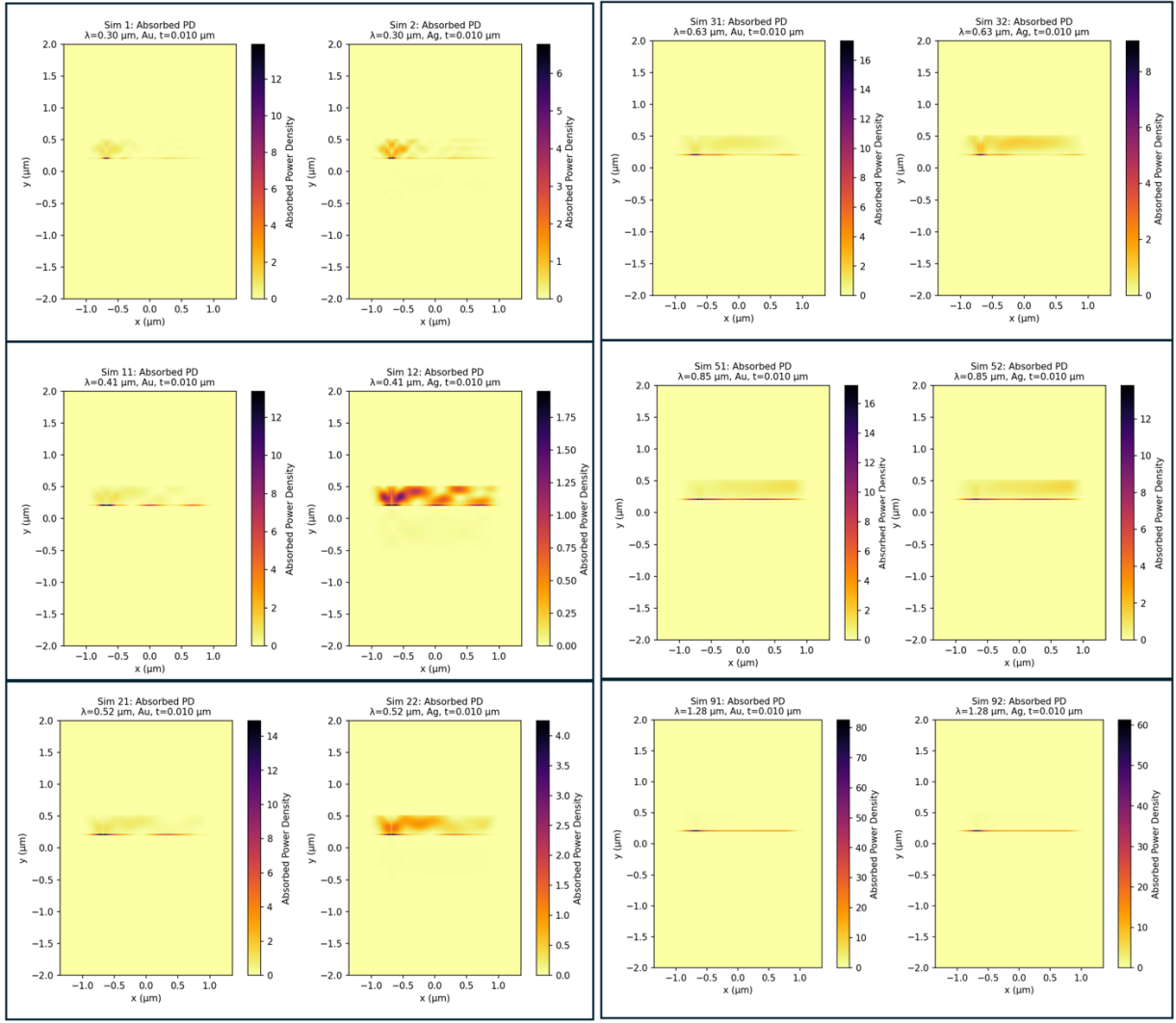

*Figure 3*: Representative FDTD-simulated spatial distributions of absorbed power density in Au and Ag plasmonic layers (thickness = 10nm) at selected wavelengths (0.30–0.85 µm). Darker (purple) regions indicate higher absorption, primarily localized near the metal–dielectric interface due to strong field confinement and interband effects below ~450 nm. At longer wavelengths (600–850 nm), the absorption hotspots become more spread out as the SPR condition shifts and eventually weakens. Au's broader absorption profile reflects its larger interband contributions, whereas Ag's lower imaginary dielectric component yields a slightly sharper but overall weaker response. These spatial maps corroborate the overall trends in *Figure 2*, illustrating how material properties, wavelength, and film thickness collectively govern the plasmonic absorption landscape.

The spatial distribution of absorption within the multilayer is illustrated in **Figure 4**, which plots the volumetric absorption density (power loss per unit volume) across a cross-section of the structure at the plasmon resonance wavelength. The absorption is highly localized at the metal–dielectric interfaces, where the surface plasmon polariton modes are concentrated. In the case of a 20 nm Au film, intense absorption occurs right at the top surface of the Au (the air/Au interface) and at the bottom interface with the dielectric spacer. These correspond to the regions of strongest electric-field intensity associated with the excited plasmon mode. By contrast, the interior of a thick metal layer sees much lower absorption, since the field decays exponentially away from the interfaces (the skin depth in Au at visible frequencies is on the order of 20 nm [19]). We also observe that the dielectric spacer plays a key role in the spatial absorption profile: the field penetrates into the spacer, and if the spacer is thin enough, the plasmon fields at the top and bottom metal interfaces can couple. This coupling gives rise to symmetric and antisymmetric plasmonic modes (often termed long-range and short-range surface plasmons in thin metal slabs [18]) and can further enhance absorption in certain regions. Overall, the simulations confirm that most of the incident energy is dissipated within a few tens of nanometers of the metal surfaces. This finding is consistent with the fundamental notion that surface plasmons are interface-bound modes with sub-wavelength field confinement [18]. From a design standpoint, it implies that adding additional absorbing layers or nanostructuring the interfaces (to create more "hot spots") could increase overall absorption by capitalizing on these intense interface fields.

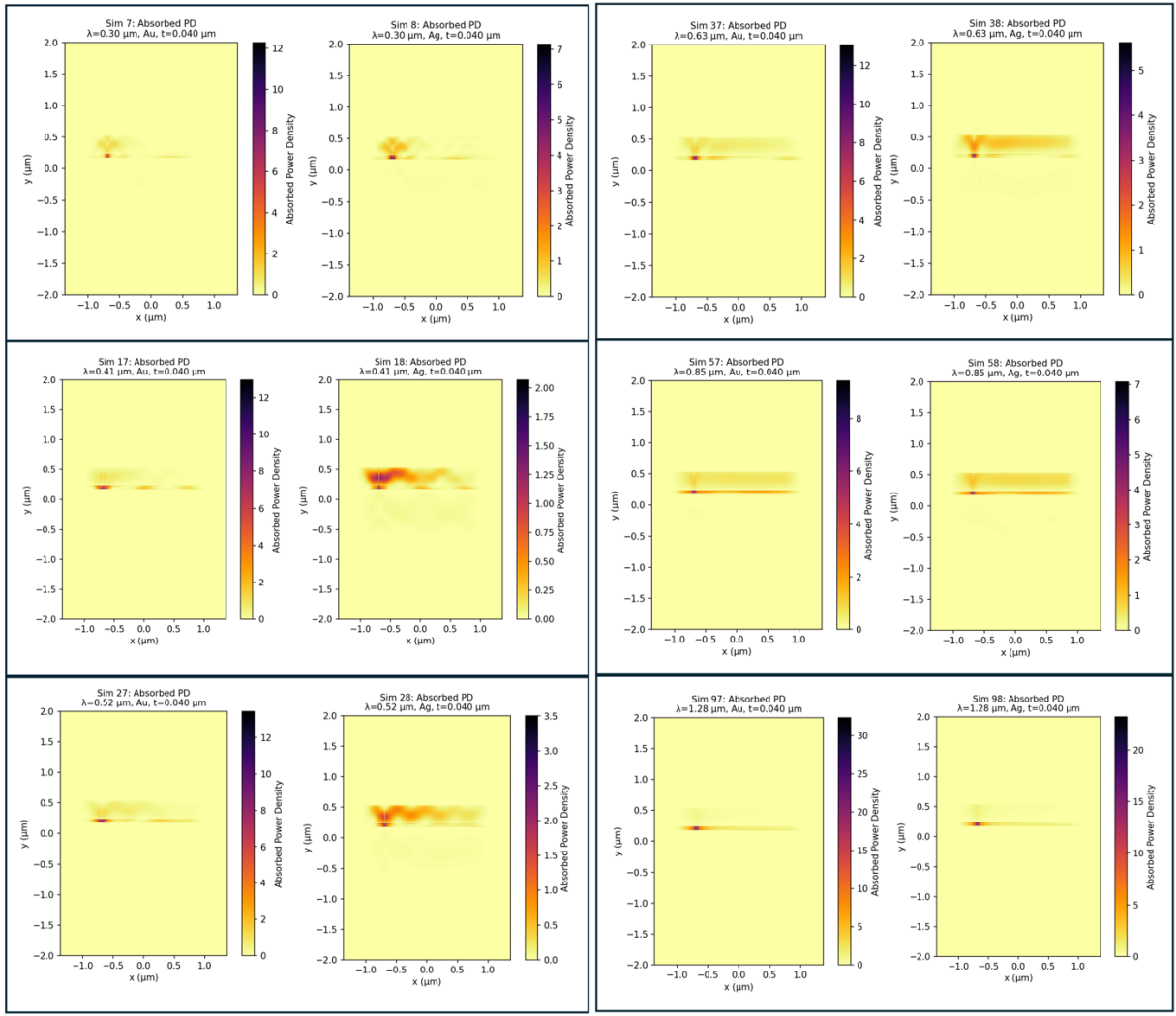

*Figure 4*: *Spatial distributions of absorbed power density (a.u) for thicker (40 nm) Au and Ag plasmonic layers, obtained via finite-difference time-domain (FDTD) simulations at representative wavelengths (0.30–1.28 µm). Higher (purple) color intensities indicate regions of elevated absorption, predominantly localized at the metal–dielectric interfaces. Compared to thinner films (**Figure 3**), the increased thickness modifies the strength and spread of these "hot spots," underscoring the influence of geometry on field confinement and absorption characteristics in the mid- to near-infrared range.*

## Extinction Coefficient and Dielectric Tuning Effects

To quantify the absorption efficiency of the multilayer, we extracted an effective extinction coefficient $k_{eff}(\lambda)$ from the simulation data. The extinction coefficient is related to the absorption coefficient α by $k = \alpha\lambda / (4\pi)$, and it represents the imaginary part of the effective refractive index of the medium (in this case, the multilayer stack). Using the simulated transmittance $T(\lambda)$ through the structure of total metal thickness d, we estimated $\alpha(\lambda) \approx -\ln[T(\lambda)] / d$ (assuming minimal reflectance at resonance). **Figure 5** plots the resulting $k_{eff}(\lambda)$ for selected Au layer thicknesses. We find that at the plasmon resonance (~600 nm), $k_{eff}$ increases with Au thickness and begins to approach the bulk Au value for films thicker than ~40 nm. For example, at 600 nm, the effective k for a 5 nm Au layer is around 0.5, whereas for 40 nm it reaches about 1.5, compared to the bulk gold value $k_{bulk} \approx 1.8$ at that wavelength [19]. This trend reflects the fact that thicker films absorb more strongly (higher optical density), converging to bulk-like behavior in the limit of thick layers. At the same time, the spectral dependence of $k_{eff}$ retains clear signatures of the plasmonic resonance. All thicknesses show a peak in $k_{eff}$ at the SPR wavelength, and a dip at longer wavelengths where the metal becomes more reflective (low loss). At shorter wavelengths (below ~500 nm), $k_{eff}$ rises again due to the onset of interband transitions in Au, which contribute additional absorption independent of the plasmon mode. These features are in good agreement with ellipsometric measurements of continuous Au films, indicating that our multilayer behaves as an effective medium with optical constants similar to those of homogeneous metal, modified by interference effects.

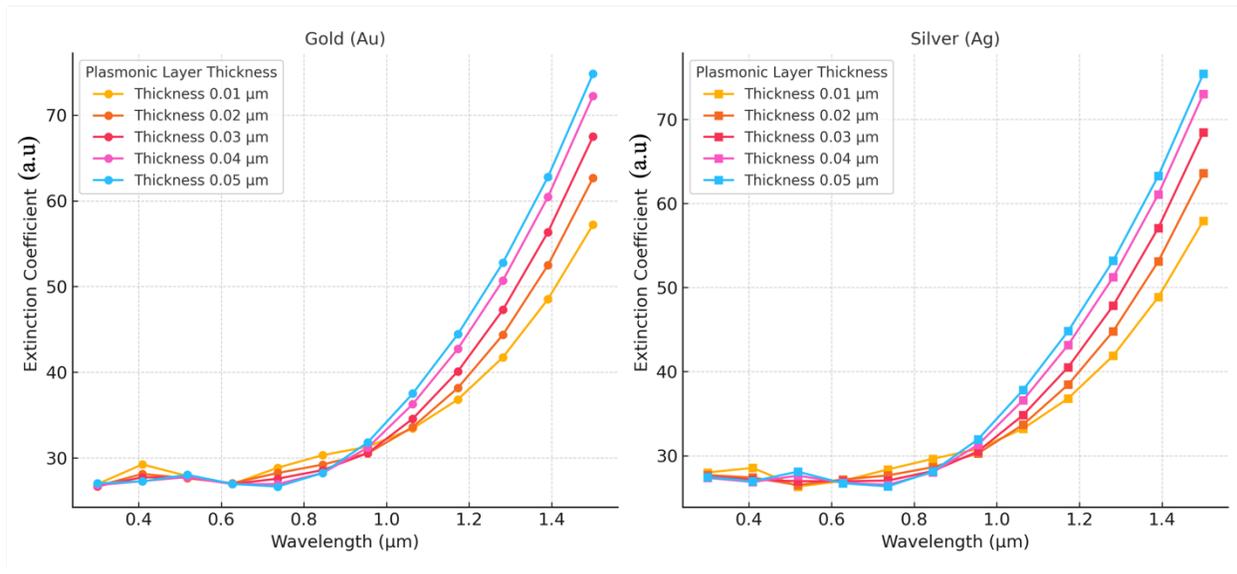

*Figure 5*: Extinction coefficient as a function of wavelength for Au (left) and Ag (right) plasmonic layers with varying thicknesses. The extinction coefficient is calculated for each material and layer configuration, with markers distinguishing different layer thicknesses ranging from 10 μm to 50 nm. As wavelength increases, the extinction coefficient rises. Differences between Au and Ag reflect the materials' distinct optical responses and damping rates, with thicker layers exhibiting higher extinction, consistent with the confinement and dissipation of the electromagnetic field within the plasmonic layer.

In addition to layer thickness, the dielectric properties of the surrounding media provide another lever to tune the optical response. We performed simulations with different refractive indices for the spacer and capping layers to examine how changes in the dielectric environment affect the plasmonic absorption. As expected, increasing the refractive index *n* leads to a red-shift of the absorption peak along with a moderate increase in its magnitude. For instance, when the spacer index is increased from n = 1.0 (air) to n = 1.5, the SPR wavelength shifts from about 580 nm to 620 nm, and the peak absorption rises slightly. The red-shift occurs because a higher $\varepsilon_d$ in the surrounding medium requires a lower oscillation frequency (longer wavelength) to satisfy the resonance condition $Re[\varepsilon(\omega)] \approx -\varepsilon_d$ [1]. The increase in absorption intensity can be attributed to improved coupling, a higher-index environment tends to better confine the field and reduce radiation losses, effectively increasing the mode overlap with the metal. The tuning rate we observe, on the order of tens of nanometers shift in resonance per 0.1 change in index, is consistent with reported sensitivities of surface plasmon sensors based on thin metal films [17]. We note that in some configurations (e.g., if a metallic back-reflector is present), increasing the index could instead reduce absorption by detuning an optical cavity; however, in our case of a single-pass absorbing stack, the overall effect of a higher ambient index is to enhance absorption and shift it to longer wavelengths. This dielectric tuning provides a convenient mechanism for modulating the multilayer's absorption spectrum without altering its physical dimensions. In practical terms, one could exploit this by incorporating an electro-optic material or a phase-change material as the spacer, thereby actively controlling *n* (and hence the plasmon resonance) via an external stimulus. Such tunable plasmonic devices would be interesting for reconfigurable filters and modulators.

**Machine Learning Prediction of Optical Responses**

We employed ML models to rapidly predict the optical response of the multilayer across a broad design parameter space. A feed-forward MLP and a CNN were trained on simulation data consisting of absorbed power and flux values generated from FDTD simulations across numerous combinations of layer thicknesses (10–50 nm), material composition (Au or Ag), and excitation wavelengths (300–1500 nm). The input to each model encoded the structural parameters, namely the thickness of the plasmonic layer, the type of metal, and the dielectric configuration, while the output corresponded to either the integrated absorbed power and flux (MLP) or the spatial distribution of absorbed power density (CNN). Both models demonstrated high predictive accuracy, with MAE below 2% on test data, validating their ability to emulate full-wave simulations with minimal loss of fidelity. **Figure 6** illustrates the training and validation loss history for the MLP, showing steady convergence and no evidence of overfitting. The resulting predictions of absorbed power and flux closely track the simulation data, as shown in **Figure 7**, where parity plots of predicted versus actual values reveal a strong linear correlation and minimal scatter. The CNN model outperformed the MLP in spatial accuracy, owing to its convolutional architecture's ability to capture local spatial correlations and interlayer interactions. As illustrated in **Figure 8**, the CNN accurately reproduces absorbed power density maps across the multilayer structure, preserving key spatial features such as plasmonic "hot spots" near metal–dielectric interfaces and the expected decay profiles through thicker metallic regions. These spatial maps validate the CNN's ability not only to infer global metrics but also to approximate field distributions, a critical capability for evaluating near-field enhancement and device functionality.

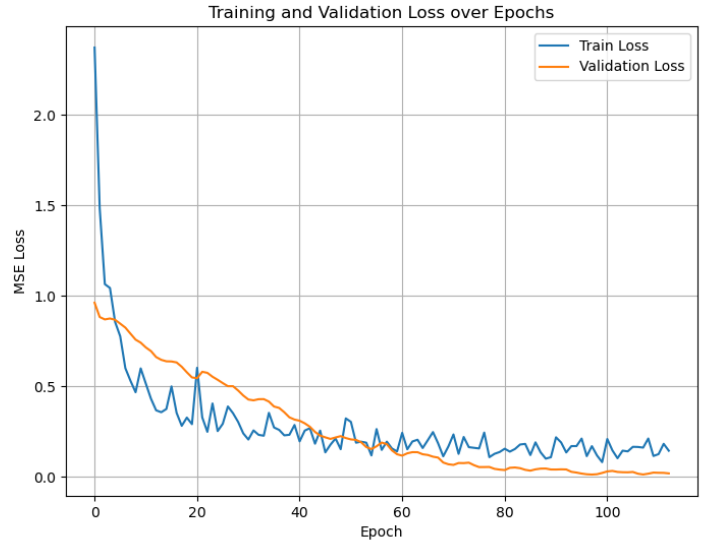

*Figure 6*: *Train and Validation loss as a function of the number of Epochs in the MLP model*

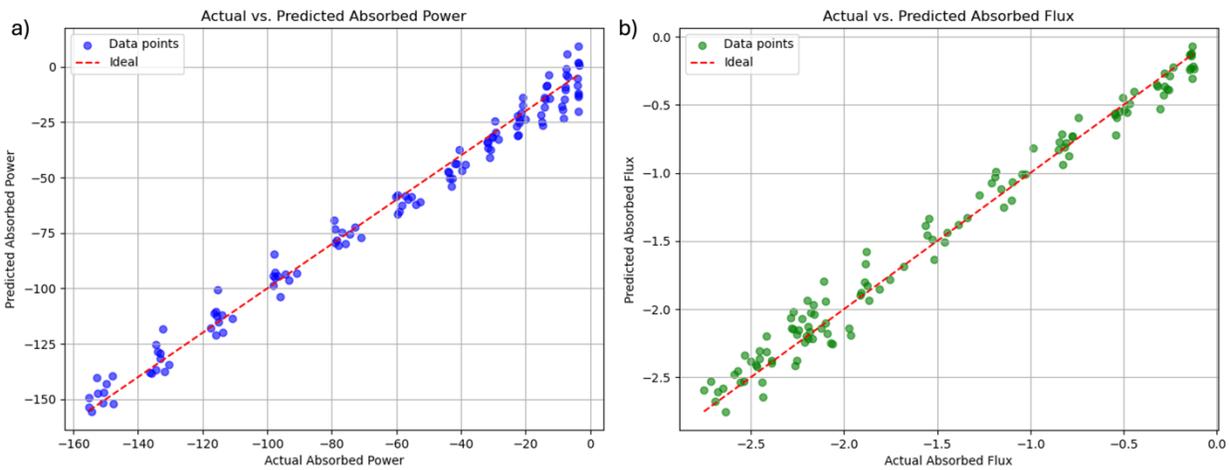

*Figure 7*: *Comparison of predicted and actual absorbed values for power and flux. (a) Scatter plot of actual versus predicted absorbed power, with the red dashed line representing the ideal 1:1 correspondence. The strong alignment of data points along the ideal line indicates accurate model predictions. (b) Scatter plot of actual versus predicted absorbed flux, also showing good agreement with the ideal prediction line. The distribution of points demonstrates the MLP's model capability to capture flux variations effectively, with minimal deviation.*

Our application of neural networks in this context builds upon the growing integration of machine learning in nanophotonics and plasmonics[11,21–23]. For example, Peurifoy et al. used neural networks to model and inversely design multilayer nanoparticles, while Malkiel et al. demonstrated CNNs for metasurface optimization [21,22]. Yeung et al. further introduced explainable AI for nanostructure behavior interpretation, and Liu et al. provided a broader overview of deep learning across photonic design domains [11,23].

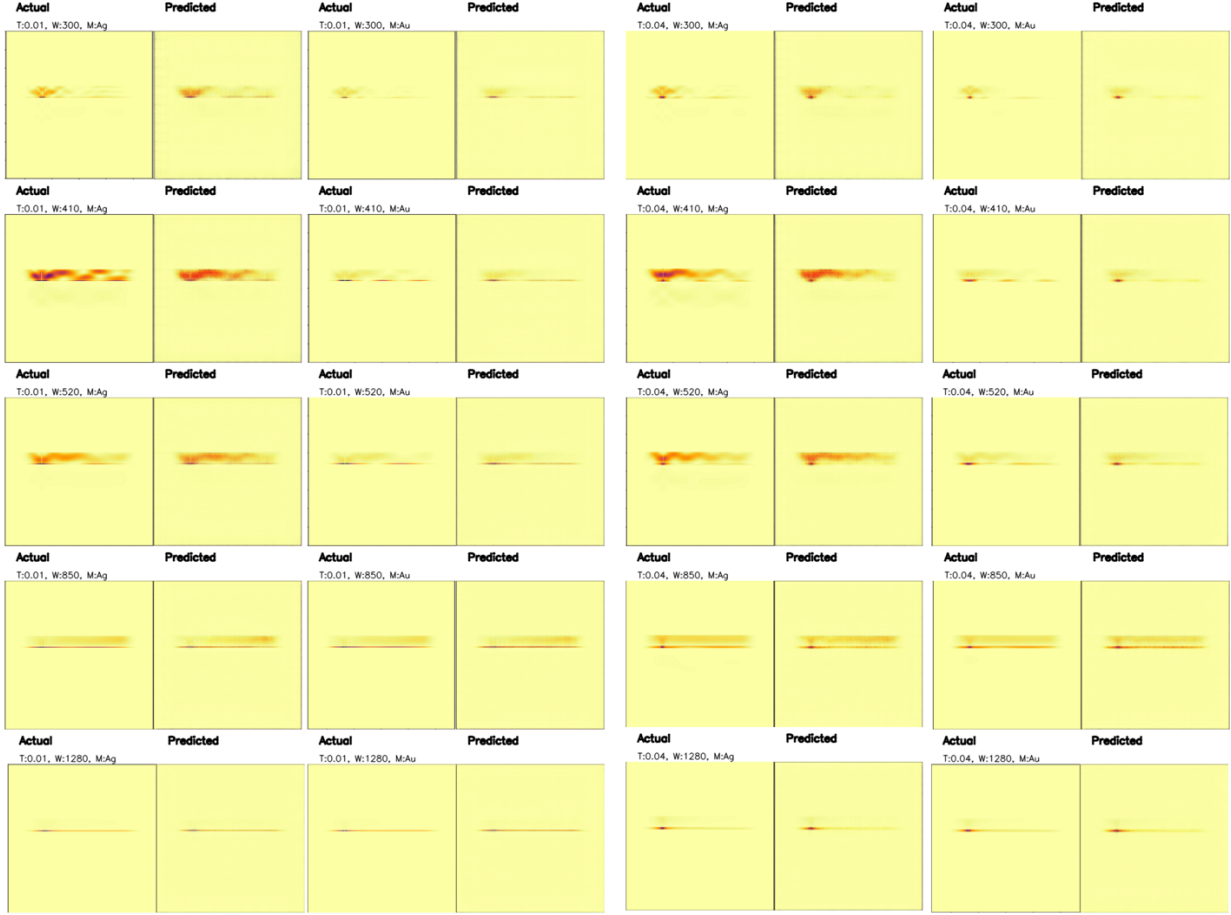

*Figure 8*: *Actual and predicted spatial maps of absorbed power density on a 4 × 2.7 system with a 500 nm thick SiO₂ substrate. The plasmonic layers (M) consist of either Au or Ag with thicknesses (T) varied from 0.01 to 0.05 µm. The top layer is a 200 nm thick ITO layer, and the wavelength (W) is varied from 300 to 1500 nm.*

However, unlike these studies, which focused primarily on isolated nanoparticles or 2D metastructures, our work targets continuous planar multilayer stacks, an underexplored yet fundamentally important geometry in thin-film optics and plasmonics. In such systems, the absorption response results from a combination of delocalized surface plasmon polariton modes and multilayer interference, rather than localized resonances. Moreover, the thin-film stack exhibits strong sensitivity to nanometer-scale thickness changes, dielectric environment, and excitation wavelength, dependencies our models successfully learn and generalize across. While prior ML frameworks have focused on spectral outputs alone [21,22], our CNN model also predicts spatial power density distributions, enabling near-field analysis and guiding field enhancement engineering. This dual capability provides a more complete toolset for the inverse design of multilayer devices such as broadband absorbers, sensors, and modulators.

By training on high-fidelity simulation data and validating against known plasmonic trends [17–19], the models not only achieve computational acceleration over traditional solvers but also produce interpretable results, capturing known physical behaviors such as absorption saturation with increasing metal thickness and red-shifted plasmonic resonances with increasing dielectric index. In this regard, our framework offers a bridge between data-driven prediction and physics-based design, opening new pathways for efficient exploration and tailored engineering of complex photonic thin films.

## SHAP Analysis of Feature Importance

While the predictive accuracy of our ML models is high, it is also important to extract physical insights from these models. We therefore applied SHAP analysis to interpret the trained MLP and CNN models. SHAP is an explainable AI technique that assigns each input feature a "Shapley value" indicating the impact of that feature on the model's output (positive or negative), based on cooperative game theory. In our context, the features are the design parameters (Au thickness, Ag thickness, spacer thickness, refractive index, etc.), and we computed SHAP values to understand how each parameter influences the predicted absorption. **Figure 9** summarizes the SHAP analysis for the CNN model, focusing on the contribution of each feature to the predicted peak absorption. We find that the Au layer thickness is the most influential parameter: large Au thickness generally drives the model to predict high absorption, until it saturates at the optimal value (beyond which further increases slightly reduce the predicted absorption, as indicated by the SHAP value changing sign). The Ag layer thickness is the second most important feature; interestingly, the SHAP plot reveals a point of diminishing returns where increasing the Ag thickness beyond ~20 nm yields little additional absorption, aligning with the physical expectation of an optimal thickness. The spacer refractive index also shows a strong positive influence on absorption in the SHAP analysis, confirming that higher index environments are predicted to enhance coupling and absorption (up to the point of impedance mismatch). On the other hand, the spacer thickness has a more nuanced effect: intermediate spacer thickness (on the order of half the resonance wavelength) is favorable for forming a resonant cavity, but excessively thick spacers decouple the metal layers and reduce absorption, which is reflected by negative SHAP contributions for large spacer thickness. These interpretations from the ML model comport well with our direct simulations and theoretical understanding, giving us confidence that the network has learned meaningful physics rather than spurious correlations.

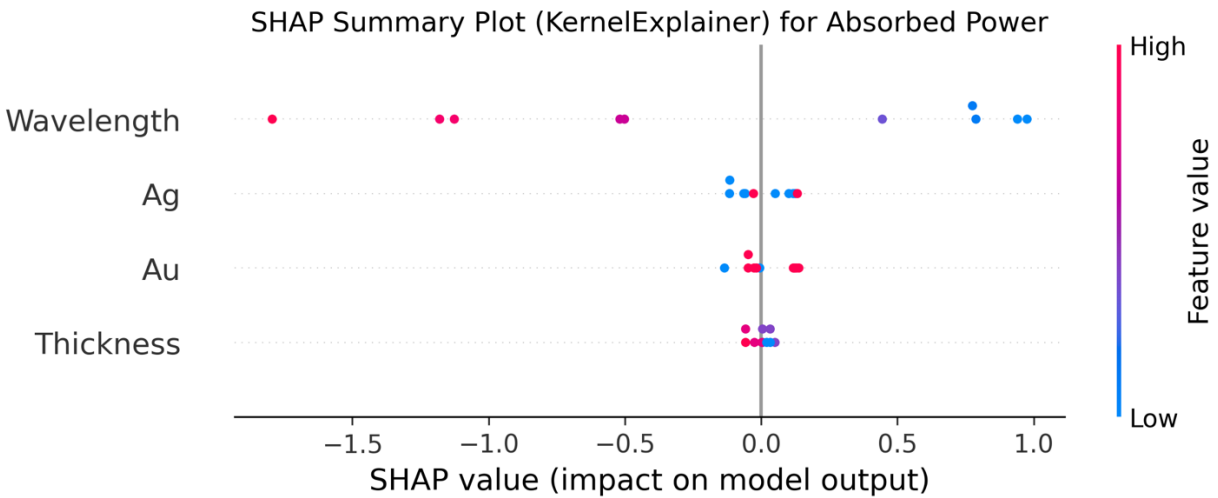

*Figure 9*: SHAP (SHapley Additive exPlanations) summary plot illustrating the relative influence of key input features—wavelength, thickness, and metal type (Au or Ag) – on the MLP's predicted absorbed power. Each dot represents one simulation sample, color-coded by feature value (blue indicates lower values, red higher). The horizontal axis denotes the SHAP value, with positive values signifying an increase in predicted absorption and negative values indicating a decrease. The pronounced spread in wavelength's SHAP values underscores its dominant role, consistent with earlier resonance-based findings, while variations in film thickness and material composition further fine-tune absorption behavior.

In addition to global feature importance, we used SHAP values to explain individual predictions and identify design outliers. For example, for a particular test sample where the CNN predicted a relatively low absorption, the SHAP analysis indicated that the sample had an unusually thin Au layer and low-index spacer – a combination that we know is suboptimal for plasmonic absorption, thus explaining the model's prediction. This ability to interpret the ML predictions is crucial if such models are to be used in the design process, as it provides guidance on *why* a certain design performs poorly or well and how to improve it. The concept of explainable machine learning in photonics is relatively new, but initial studies have demonstrated its value in revealing the connections between geometry and optical response that a neural network has learned. Yeung *et al.* recently showed that a CNN trained on the spectral responses of metamaterials could be interrogated to identify which sub-regions of the structure were most responsible for particular optical features, thus bridging the gap between the black-box model and physical intuition [23]. In our planar multilayer case, SHAP analysis serves a similar purpose: it bridges ML predictions and human understanding by attributing outcomes to specific input factors. The consistency between the SHAP-derived importances and known physical trends (e.g., metal thickness being the dominant factor, dielectric index tuning the resonance, etc.) not only validates our model but also provides a quantitative ranking of design parameters by their efficacy. Such insights could inform experimental efforts by highlighting which parameters deserve tight control or active tuning. In summary, incorporating explainable AI techniques like SHAP allows us to treat the ML model as an aid for discovery, using it to gain new perspectives on the device physics and to accelerate the optimization of plasmonic structures.

CONCLUSION

In summary, we have conducted a comprehensive study of the optical properties of plasmonic Au/Ag multilayer films, combining electromagnetic simulations, theoretical modeling, and machine learning analysis to understand their behavior. The results show that these multilayer structures support strong wavelength-dependent absorption resonances arising from plasmonic excitations, and that the spectral position and intensity of these resonances can be tuned by adjusting the layer thicknesses and the dielectric environment. Using the Drude–Lorentz theory, we explained the observed resonance features in terms of the metal's dielectric function: the free-electron plasma oscillation in the metal (with a plasma frequency in the ultraviolet) gives rise to a negative permittivity and a resultant SPR in the visible range, while interband transitions in Au (above ~2.4 eV) contribute additional absorption that broadens the resonance on the short-wavelength side. We quantified the effect of metal thickness on absorption, identifying an optimal thickness that maximizes plasmonic absorption and noting the diminished response of ultra-thin (sub-5 nm) layers due to incomplete film percolation. Spatial absorption profiles from simulations confirmed that absorption is highly localized at the metal–dielectric interfaces, consistent with the notion that surface plasmons confine electromagnetic energy to the nanoscale vicinity of the interfaces. We also demonstrated that the dielectric spacer index can be used to tune the plasmon resonance wavelength over tens of nanometers, providing a practical knob for spectral control without changing physical dimensions.
Furthermore, by leveraging machine learning, we developed surrogate models (MLP and CNN) that can predict the multilayer's optical response with near-physical accuracy. The CNN model, in particular, was able to capture the complex dependence of the absorption spectrum on multiple coupled parameters, outperforming the simpler MLP and achieving <2% error. This data-driven approach enables instant evaluation of new designs, dramatically speeding up the exploration of the design space. Importantly, through SHAP analysis, we extracted interpretable insights from the ML models, confirming that the networks' predictions were based on physically sensible trends

(e.g., recognizing the critical importance of metal thickness and dielectric properties). The interpretability not only builds trust in the model but also provides guidance for design: for instance, the ML analysis suggests focusing on optimizing the gold thickness and the refractive index contrast to achieve maximum absorption, in agreement with our theoretical understanding. Altogether, this work illustrates a powerful synergistic approach for plasmonic device engineering, where theory, simulation, and machine learning inform each other. We have provided a detailed understanding of how plasmonic multilayers behave as a function of key parameters, and we have shown that simple neural networks can learn this behavior to facilitate rapid design iterations. These insights can be applied to the development of advanced optical coatings and absorbers for applications in sensing, spectroscopy, and energy harvesting, where controlled light absorption is critical. The strategy of using explainable ML to supplement physical intuition is broadly applicable to other nanophotonic systems as well. Indeed, the integration of machine learning techniques into nanophotonics is increasingly recognized as a means to accelerate discovery and optimize complex structures. By demonstrating both high-fidelity predictions and human-interpretable analysis, our work paves the way for an AI-assisted paradigm in photonics research, in which researchers can efficiently navigate design spaces and uncover fundamental relationships in multidimensional data. We envision that the methodologies presented here will aid in the rational design of future plasmonic multilayer devices and inspire further adoption of interpretable machine learning in photonics.

METHODS

The multi-faceted approach integrates high-fidelity finite-difference time-domain (FDTD) simulations and convolutional neural networks (CNNs) to achieve accurate predictions across broad parameter ranges.

**Simulation Setup and Theoretical Framework**

To model electromagnetic interactions within the multi-layer plasmonic structures, we employed the FDTD method using the Meep simulation package, which solves Maxwell's equations in both time and space domains. The mathematical framework is based on solving the following coupled differential equations governing the electric ($E$) and magnetic ($H$) fields:

$$\nabla \times E = -\mu_0 \frac{\partial H}{\partial t}, \quad \nabla \times H = \epsilon_0 \frac{\partial E}{\partial t} - \sigma E$$

where $\epsilon_0$ and $\mu_0$ are the vacuum permittivity and permeability, respectively, and $\sigma$ represents the electrical conductivity of the materials. The material permittivity $\epsilon(\omega)$ is frequency-dependent and modeled using:

$$\epsilon(\omega) = \epsilon'(\omega) + \epsilon''(\omega)$$

Here, $\epsilon'(\omega)$ governs dispersion, while $\epsilon''(\omega)$ accounts for absorption losses, a critical parameter for plasmonic applications. The absorbed power density $W_{abs}$ is calculated using:

$$W_{abs} = \frac{1}{2}\omega\epsilon''(\omega)|E|^2$$

The total absorbed power is obtained by integrating $W_{abs}$ over the simulation domain:

$$P_{abs} = \int_{domain} W_{abs}\, dV$$

The simulation domain comprises horizontal layers of SiO2, Au, and ITO, each assigned experimentally validated permittivity values.

**Domain and Boundary Conditions**: The 2D domain is designed to reduce computational complexity while retaining physical accuracy for layered systems. However, this simplification may limit the generality of the results when applied to 3D nanostructures where field confinement can differ significantly. For multilayer thin-film designs, 2D simulations remain effective in capturing essential absorption characteristics, but care should be taken when extrapolating to non-planar geometries. The computational cell is set to 4 μm by 2.75 μm with a resolution of 150 pixels/μm to capture fine spatial details. Perfectly Matched Layers (PML) of thickness 1.0 μm are applied to the domain boundaries to prevent wave reflections. No symmetries are enforced to preserve generality.

**Source Configuration**: A Gaussian source spanning 300 to 1500 nm introduces electromagnetic waves perpendicular to the layered structure. The incident light propagates from the left side of the structure.

**Data Generation and Preprocessing**

To explore the design space, simulations were conducted by varying the thickness of the gold and silver layers and the wavelengths of incident light, with each simulation capturing the spatial distribution of absorbed power density across the device. The dataset generated from these simulations serves as the training foundation for two predictive models: a feedforward neural network for absorbed power and flux predictions and a convolutional neural network (CNN) for spatial map predictions.

The parameter ranges include Au and Ag plasmonic layer thicknesses, discretized between 10 and 50 nm, and wavelengths sampled between 300 and 1500 nm to capture resonant behavior. Simulation results provide two key outputs: absorbed power and absorbed flux, along with spatial absorption maps. Numerical features, such as layer thickness and wavelength, are normalized using a StandardScaler, while material types are encoded using one-hot encoding to ensure compatibility with neural network inputs. Missing or anomalous data points are handled by imputing local averages, ensuring that the model input data remains complete and robust.

**Feedforward Neural Network Model for Absorbed Power and Flux Predictions**

The regression model uses a multi-layer perceptron to predict absorbed power and absorbed flux directly from the design parameters, such as layer thickness, wavelength, and material composition. The architecture consists of several dense layers, with 128 neurons per layer and ReLU activations. Dropout layers (0.2 dropout rate) and batch normalization are incorporated to enhance generalization and prevent overfitting.

The model is trained using the Adam optimizer with a learning rate of 0.001 and a mean squared error loss function. The dataset is partitioned into 80% for training and 20% for testing. Early stopping with a patience of 15 epochs is applied, and learning rate reduction on plateau is used to achieve optimal convergence.

**Convolutional Neural Network (CNN) Model Architecture**

The CNN architecture is specifically designed to capture the spatially correlated nature of the absorbed power density data. The model takes simulation parameters as inputs, which include thickness, wavelength, and material encodings, and reshapes the feature vector back into its spatial representation. Successive convolutional layers with ReLU activations are used to extract spatial features, while batch normalization stabilizes training. Dropout layers (0.3 rate) are used to prevent overfitting. The output layer produces the predicted spatial map of absorbed power density.

Hyperparameter optimization, guided by a Fibonacci-inspired scaling strategy, ensures a balance between underfitting and overfitting. The model is trained using an MSE loss function and validated on 15% of the dataset, with the remaining 15% reserved for testing. The training procedure involves 200 epochs with a batch size of 64.

**Training Procedure**

Both models (MLP and CNN) are trained using the Adam optimizer due to its adaptive learning rate capabilities. The loss function for both models is MSE:

$$L(\theta) = \frac{1}{N}\sum_{i=1}^{N}||y_i - f(x_i;\theta)||^2$$

where $N$ is the number of training samples, $y_i$ is the ground truth output, and $f(x_i;\theta)$ is the prediction.

Early stopping with a patience of 15 epochs is implemented to prevent overfitting. Additionally, learning rate reduction on plateau dynamically adjusts the learning rate when validation performance stagnates.

**Feature Importance and SHAP Analysis**

To interpret the predictions of the MLP, SHapley Additive exPlanations (SHAP) analysis is used. SHAP quantifies the contribution of each input feature (e.g., layer thickness, wavelength and material type) to the model outputs – absorbed power, providing valuable insights into which parameters most strongly influence absorption efficiency. This interpretability aids in experimental optimization and design refinement.

**Computational Requirements and Cost**

The dataset used for model training is derived from Finite-Difference Time-Domain simulations, performed on a cluster powered by a 2.2 GHz 6-Core Intel Core i7 MacBook Pro. The simulations required 17 hours to complete 120 cases due to the computational intensity of modeling spatially resolved power absorption. However, once trained, the MLP and CNN models reduce computational costs significantly by providing accurate predictions without requiring exhaustive parameter sweeps. Training the CNN model took less than 3 minutes on the same machine, while the regression model was trained in 110s. Prediction times for both models are in 1 sec, making them practical for real-time optimization tasks, compared to the 17 hours it took to generate 120 simulations from the FDTD model.

## AUTHOR INFORMATION

**Corresponding Author**

*ebamidele3@gatech.edu or emmanuel.bamidele@colorado.edu

**Present Addresses**
†Computer Science Department, Georgia Institute of Technology, USA

**Data Availability**

The complete dataset, source code, and trained models are publicly available on the project's GitHub repository (link: https://github.com/Emmanuel-Bamidele/FDTD-Bamidele). The repository contains instructions for reproducing the simulations, training the models, and conducting SHAP analysis.